\documentclass[aps,prb,showpacs,superscriptaddress,twocolumn]{revtex4}
\usepackage{amsmath}
\usepackage{amsfonts}
\usepackage{graphicx}
\usepackage{psfrag}

\begin{document}

\title{Quantum rings with time dependent spin-orbit coupling: \\
Rabi oscillations, spintronic Schr\"{o}dinger-cat states, and conductance properties}
\author{P\'{e}ter F\"{o}ldi}
\affiliation{Department of Theoretical Physics, University of Szeged, Tisza Lajos k\"{o}r%
\'{u}t 84, H-6720 Szeged, Hungary}
\author{Mih\'{a}ly G. Benedict}
\email{benedict@physx.u-szeged.hu}
\affiliation{Department of Theoretical Physics, University of Szeged, Tisza Lajos k\"{o}r%
\'{u}t 84, H-6720 Szeged, Hungary}
\author{Orsolya K\'{a}lm\'{a}n}
\affiliation{Department of Quantum Optics and Quantum Information, Research Institute for Solid
State Physics and Optics,Hungarian Academy of Sciences, Konkoly-Thege Mikl%
\'{o}s \'{u}t 29-33, H-1121 Budapest, Hungary}
\author{F. M. Peeters}
\email{francois.peeters@ua.ac.be}
\affiliation{Departement Fysica, Universiteit Antwerpen, Groenenborgerlaan 171, B-2020
Antwerpen, Belgium}

\begin{abstract}
The strength of the (Rashba-type) spin-orbit coupling in mesoscopic
semiconductor rings can be tuned with external gate voltages. Here we
consider the case of a periodically changing spin-orbit interaction strength as
induced by sinusoidal voltages. In a closed one dimensional quantum
ring with weak spin-orbit coupling, Rabi oscillations are shown to appear.
We find that the time evolution of initially localized wave
packets exhibits a series of collapse and revival phenomena. Partial
revivals -- that are typical in nonlinear systems -- are shown to correspond
to superpositions of states localized at different spatial positions along
the ring. These "spintronic Schr\"odinger-cat sates" appear periodically, and similarly to their counterparts in other physical systems, they are found to be sensitive to environment induced disturbances.
The time dependent spin transport problem, when leads are attached to the ring,
is also solved. We show that the "sideband currents" induced by the oscillating spin-orbit interaction strength can become the dominant output channel, even in the presence of moderate thermal fluctuations and random scattering events.\end{abstract}

\maketitle

\section{Introduction}

The observation of fundamental flux- and spin-dependent quantum interference phenomena that
can appear in quantum rings made of semiconducting materials \cite{KTHS06}
exhibiting Rashba-type\cite{R60} spin-orbit interaction\cite%
{G00,NATE97,SKY01} (SOI) have motivated many studies in the past few years.
Additionally, as the strength of the SOI that determines the spin
sensitive behavior can be tuned with external gate voltages\cite{NATE97},
quantum rings or systems of them\cite{BKSN06,ZW07,KFBP08b,KFBP08c,CSz08} also have possible spintronic applications.
Because of the conceptual importance and the possible applications, closed
single quantum rings (without attached leads),~\cite{SGZ03,SC06,Y06,YCSX07b,CPC08}
as well as two- or three-terminal ones were investigated \cite{BIA84,X92,NMT99,YPS03,FHR01,MPV04,FR04,FMBP05,WV05,SzP05,SN05,KMGA05,SP05,VKN06,BO07,VKPB07,CHR07,BO08,WC08,CPC08}
extensively.

In the current paper we explore effects which are
expected to appear when the SOI in a single quantum ring is time dependent.
The strength of this interaction is assumed to be a sinusoidally oscillating
function of time. Studies of transport related problems with oscillating SOI
have been initiated in Ref.~[\onlinecite{WC07}] for a ring, and in Ref.~[\onlinecite{RCM08}]
for a ring-dot system, mainly in the context of spin currents.
Here we focus on different aspects by starting the analysis with the case of closed
rings and determine the energy levels and eigenstates of the relevant
Hamiltonian\cite{Meijer,MPV}. Using these analytic results we can calculate
the dynamics for arbitrary initial conditions. The time evolution of states with
well-defined total angular momentum component $j$ in the direction
perpendicular to the plane of the ring is shown to be analogous to the
classical Rabi flopping. That is, the spin components oscillate with the
Rabi frequency, which in our case depends on the eigenvalue $j,$ and the oscillation itself is relatively stable against disturbances caused by random scatterers. When the
initial state is a localized wave packet, then, theoretically, the time evolution is periodic, if the ratios of the relevant eigenfrequencies are rational. We show that appropriately chosen amplitude of the SOI oscillations
leads to quasi-periodic dynamics, even when the relevant frequencies are not exactly commensurable.

The description of the conductance properties of the ring requires the solution
of a quantum mechanical scattering problem with a time dependent
Hamiltonian\cite{WC07,RCM08}. In this case energy is clearly not a constant of motion, the
relevant continuity equation contains a source term describing the energy
explicitly calculate this source term as well as the energy current density.
These results provide a clear physical picture from which we can solve the scattering
problem. As an application, we show that the ring can shift the energy of
the incoming plane waves (both up- and downwards) by the frequency of the
SOI oscillations expressed in energy units. This effect means the emergence of "sideband" currents in the transmission, which, in our case,  can become stronger than the direct one.  Clearly, harmonics of the driving frequency (which corresponds to the sideband currents in our case) appear naturally in driven nonlinear systems. However, let us emphasize that quantum rings are special in the sense that their geometry induces nontrivial interference effects that determine the energy dependent transmission probabilities. This characteristics of the device is responsible for the increased relative weight of the sideband currents in the output.

\section{Spin oscillations in a closed ring}

\subsection{Model}

We consider a ring \cite{ALG93} of radius $a$ in the $x-y$ plane and assume
a time dependent electric field in the $z$ direction controlling the
strength of the spin-orbit interaction characterized by the parameter $%
\alpha $ \cite{NATE97}. The Hamiltonian \cite{Meijer,MPV} in the presence of
spin-orbit interaction for a charged particle of effective mass $m^{\ast}$
is given by
\begin{equation}
H=\hbar\Omega\left[ \left( -i\frac{\partial}{\partial\varphi}+\frac {%
\omega(t)}{2\Omega}\sigma_{r}\right) ^{2}-\frac{\omega^{2}(t)}{4\Omega^{2}}%
\right]  \label{Ham}
\end{equation}
where $\varphi$ is the azimuthal angle of a point on the ring (see Fig.~\ref%
{ringfig}) and the radial spin operator is given by $\sigma
_{r}/2=(\sigma_{x}\cos\varphi+\sigma_{y}\sin\varphi)/2$. Additionally, $%
\hbar\Omega=\hbar^{2}/2m^{\ast}a^{2}$ denotes the kinetic energy of the
charged particle and
\begin{equation}
\omega(t)=\alpha(t)/\hbar a=2A\cos(\nu t),  \label{omt}
\end{equation}
with $\nu$ being the circular frequency of the periodic external electric
field. Nanoscale quantum rings, for which the Hamiltonian above is relevant,
can be fabricated from e.g.~InAlAs/InGaAs based heterostructures\cite{KNAT02} or
HgTe/HgCdTe quantum wells\cite{KTHS06}. For a ring of radius 250 nm made of
InGaAs, the frequency corresponding to $\Omega$ is around $10$ GHz.
\begin{figure}[tbh]
\psfrag{in}{$|\Psi_I\rangle$} \psfrag{out}{$|\Psi_{II}\rangle$} %
\psfrag{up}{$|\Psi_{u}\rangle$} \psfrag{down}{$|\Psi_{l}\rangle$} %
\psfrag{phi1}{$\varphi_1$} \psfrag{phi2}{$\varphi_2$} %
\includegraphics[width=8cm]{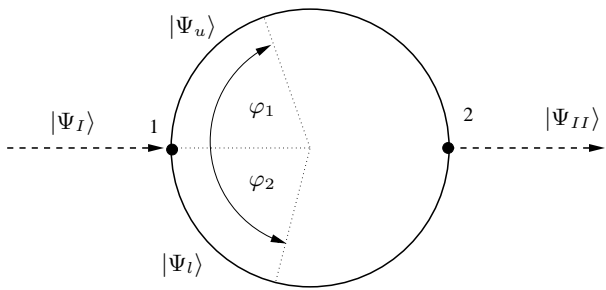}
\caption{The geometry of the device and the relevant spinor valued wave
functions in the different domains.}
\label{ringfig}
\end{figure}

In a closed ring, when there are no leads attached to it, the spinor valued
wave functions have to be periodic in space. Thus we look for the
solution of the time evolution induced by the Hamiltonian (\ref{Ham}) in
terms of states which are eigenvectors belonging to an integer eigenvalue $n$
of the $z$ component of the orbital angular momentum $L_{z}=-i\partial
_{\varphi },$ and also eigenvectors of the spin component $S_{z}=\sigma
_{z}/2$ (in units of $\hbar ).$ These operators, however, do not commute
separately with $H$, whereas in each fixed moment of time we have the commutator $%
\left[ H,L_{z}+S_{z}\right] =0,$ therefore an eigenvalue of $J=$ $%
L_{z}+S_{z} $ is a good quantum number \cite{FMBP05a}. A given eigenvalue $j$ of $%
J$ is, however, still doubly degenerate, the two space-dependent spinors
\begin{equation*}
\left\vert \uparrow ,n\right\rangle =%
\begin{pmatrix}
e^{in\varphi } \\
0%
\end{pmatrix}%
,\ \ \left\vert \downarrow ,n+1\right\rangle =%
\begin{pmatrix}
0 \\
e^{i(n+1)\varphi }%
\end{pmatrix}%
\end{equation*}%
belong to the same: $j=n+1/2$ \ half integer eigenvalue of $J.$ As a
consequence -- similarly to the case of constant $\omega $ -- the eigenvalue
equation of $H$ expanded in the $\left\{ \left\vert \uparrow ,n\right\rangle
,\left\vert \downarrow ,n+1\right\rangle \right\} $ manifold reduces to
separate $2\times 2$ matrix problems corresponding to given values of $\ j.$
Restricting $H$ to one such subspace we obtain the two-dimensional matrix
\begin{equation*}
H_{j}=\hbar  \left(
\begin{matrix}
\Omega(j-1/2)^{2} & j\omega (t) \\
j\omega (t) & \Omega(j+1/2)^{2}%
\end{matrix}%
\right) .
\end{equation*}%
In order to solve the time dependent Schr\"{o}dinger equation with the
periodic coupling of the form (\ref{omt}) we look for the solution in a
subspace fixed by a given value of $j$ in the form
\begin{equation}
\left\vert \psi (t)\right\rangle =a_{j}(t)e^{i\nu t/2}\left\vert \uparrow
,j-1/2\right\rangle +b_{j}(t)e^{-i\nu t/2}\left\vert \downarrow,
j+1/2\right\rangle.
\label{intpict}
\end{equation}%
This is useful when -- as in the current paper -- the focus is on relatively
weak spin-orbit strengths, and one can use standard rotating wave
approximation (RWA), i.e., terms oscillating rapidly, with frequency $2\nu $
in the off-diagonal terms in the Hamiltonian can be neglected. Note that this
widely used approximation is essentially equivalent here to the first order corrections
(with respect to $\nu$) in an appropriate Floquet scattering matrix description\cite{LR99,MB04} of the problem.
Using RWA, and introducing the dimensionless time variable $\tau =\Omega t,$ the system of
equations determining the evolution has the form:
\begin{equation}
i\frac{d}{d\tau }\left(
\begin{array}{c}
a_{j} \\
b_{j}%
\end{array}%
\right) =\tilde{H}_{j}\left(
\begin{array}{c}
a_{j} \\
b_{j}%
\end{array}%
\right)  \label{HnS}
\end{equation}%
\ with the dimensionless operator
\begin{equation}
\tilde{H}_{j}=(j^{2}+\frac{1}{4})\boldsymbol{1}+\left(
\begin{matrix}
-(j-\frac{\nu }{2\Omega }) & \frac{Aj}{\Omega} \\
\frac{Aj}{\Omega} & (j-\frac{\nu }{2\Omega })%
\end{matrix}%
\right) .  \label{Hn}
\end{equation}

The solution of the system (\ref{HnS}) amounts to solve the eigenvalue
equation of $\tilde{H}_{j}$ yielding
\begin{equation}
E_{j}^{\pm }=(j^{2}+\frac{1}{4})\pm \frac{1}{2}\sqrt{\left( 2j-\tilde{\nu}%
\right) ^{2}+4\tilde{A}^{2}j^{2}},  \label{energies}
\end{equation}%
where $\tilde{\nu}=\nu /\Omega , \tilde{A}=A/\Omega.$ The corresponding eigenspinors are given
by
\begin{equation}
\left\vert \psi _{j}^{+}\right\rangle =%
\begin{pmatrix}
u_{j} \\
v_{j}%
\end{pmatrix}%
,\ \left\vert \psi _{j}^{-}\right\rangle =%
\begin{pmatrix}
-v_{j} \\
u_{j}%
\end{pmatrix}%
,  \label{eigs}
\end{equation}%
where $u_{j}=\cos \theta _{j}/2,$ $v_{j}=\sin \theta _{j}/2,$
$\tan \theta _{j}=\frac{2\tilde{A}j}{\tilde{\nu}-2j}$
and
\begin{equation}
\frac{v_{j}}{uj}=\frac{\sqrt{\left( 2j-\tilde{\nu}\right) ^{2}+4\tilde{A}^{2}j^{2}}%
+2j-\tilde{\nu}}{2\tilde{A}j}.
\end{equation}

Note that for a given $j,$ the spin operator $S(\theta _{j},\varphi
)=S_{x}\sin \theta _{j}\cos \varphi +S_{y}\sin \theta _{j}\sin \varphi
+S_{z}\cos \theta _{j}$ (the spin component in the direction given by $%
\theta _{j}$ and $\varphi $) also commutes with $H,$ therefore in the case
of a given $j$ the expectation value of the spin direction varies along the
ring in accordance to the $\varphi $ dependence of the operator above.

The states $\left\{ \left\vert \psi _{j}^{\pm }\right\rangle ,j=\ldots
-3/2,-1/2,1/2,3/2\ldots \right\} $ obviously form a basis in the space of
periodic spinor valued wave functions on the ring. Therefore, the time
evolution (in the interaction picture introduced above) of any initial state
\begin{equation}
\left\vert \psi (0)\right\rangle =%
\begin{pmatrix}
f(\varphi ) \\
g(\varphi )%
\end{pmatrix}%
\end{equation}%
can be obtained in a straightforward way:
\begin{eqnarray}
\left\vert \psi (\tau )\right\rangle &=&\sum_{j} \left( e^{-iE_{j}^{+}\tau
}\left\vert \psi _{j}^{+}\right\rangle \left\langle \left\langle \psi
_{j}^{+}\left\vert {}\right. \psi (0)\right\rangle \right\rangle \right. \\
&+&\left.e^{-iE_{j}^{-}\tau }\left\vert \psi _{j}^{-}\right\rangle \left\langle
\left\langle \psi _{j}^{-}\left\vert {}\right. \psi (0)\right\rangle
\right\rangle\right) ,  \notag
\end{eqnarray}%
where the inner products $\langle \langle .|.\rangle \rangle $ that provide
the expansion coefficients are the following:
\begin{eqnarray}
\left\langle \left\langle \psi _{j}^{+}\left\vert {}\right. \psi
(0)\right\rangle \right\rangle &=&u_{j}^{\ast }\frac{1}{2\pi }\int_{0}^{2\pi
}e^{-i(j-1/2)\varphi }f(\varphi )d\varphi \notag
\\
&+&v_{j}^{\ast }\frac{1}{2\pi }\int_{0}^{2\pi }e^{-i(j+1/2)\varphi
}g(\varphi )d\varphi .  \label{completeinner}
\end{eqnarray}

\subsection{State evolution of electrons with definite angular momentum:
Rabi oscillations, collapse and revival}

States with well defined $z$ component of the total angular momentum are
linear combinations of the spinors $\left\vert \psi _{j}^{+}\right\rangle $
and $\left\vert \psi _{j}^{-}\right\rangle $ with the same value of $j.$ The
time evolution of these states can be of interest because conservation of
the angular momentum may provide a method for preparing them. Absorption of
circularly polarized photons e.g.~can excite these states. (Interaction of short
light pulses with the electrons confined in a ring has been discussed in Refs.~[\onlinecite{MB05,ZB08}].)

\begin{figure}[tbh]
\psfrag{label0}{$\frac{\Omega_R \tau}{2\pi}=0$} \psfrag{label10}{$\frac{\Omega_R \tau}{2\pi}=0.05$} \psfrag{label20}{$\frac{\Omega_R \tau}{2\pi}=0.1$} %
\psfrag{label30}{$\frac{\Omega_R \tau}{2\pi}=0.15$} \psfrag{label90}{$\frac{\Omega_R \tau}{2%
\pi}=0.45$} \psfrag{label105}{$\frac{\Omega_R \tau}{2\pi}=0.5$} %
\psfrag{label157}{$\frac{\Omega_R \tau}{2\pi}=0.75$} \psfrag{label210}{$%
\frac{\Omega_R \tau}{2\pi}=1.0$} \includegraphics[width=8cm]{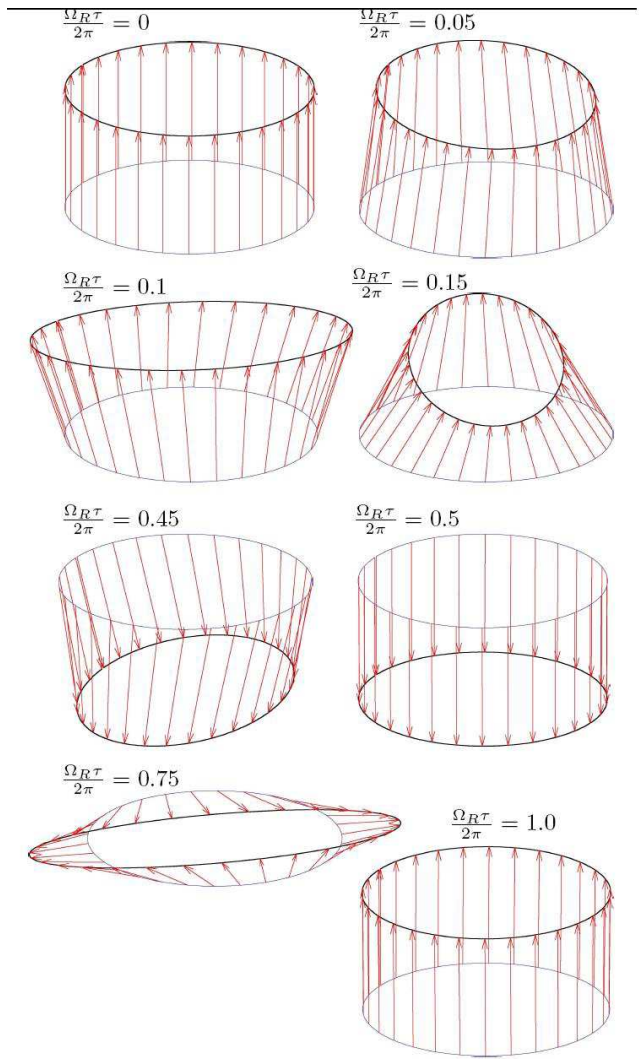}
\caption{Resonant Rabi oscillation in the quantum ring. The initial wave
function is given by Eq.~(\protect\ref{initstate}) with $j=3/2,$ while $\tilde{A}=0.1,$
leading to $\Omega _{R}=0.3.$ See EPAPS Document No. [number will be inserted by publisher] for the related movie file.}
\label{rabifig}
\end{figure}

As it is known\cite{S94}, if a two-level system is getting excited with a resonant
external field, the interaction produces Rabi oscillations of the level
populations with a frequency proportional to the amplitude of the external
field. If there is a detuning between the external field and the level
splitting, the Rabi frequency is modified and the oscillations become less
pronounced. The dimensionless Rabi frequency corresponding to the Hamiltonian $\tilde{H}%
_{j}$ (see Eq.~(\ref{Hn})), is given by
\begin{equation}
\Omega _{R}=\sqrt{\left( 2j-\tilde{\nu}\right) ^{2}+4\tilde{A}^{2}j^{2}}.
\end{equation}%
In our case the (dimensionless) level splitting $\Delta E=2j,$ the detuning $%
2j-\tilde{\nu}$, as well as the effective coupling $\tilde{A}j$ depend on $j,$ that is,
on the $z$ component of the total angular momentum. This implies that the
resonance frequency of the external field is also different for different
values of $j,$ namely $\nu _{r}=\tilde{\nu}_{r}\Omega =2j\Omega .$ When this
resonance condition is met, the weights of the spin up and spin down
components in the eigenspinors given by Eq.~(\ref{eigs}) are equal, $\tan
\theta _{j}/2=1$ (in this case independently from $j.$) Assuming that the
initial state is spin-polarized in the positive $z$ direction,
\begin{equation}
\left\vert \psi (0)\right\rangle =%
\begin{pmatrix}
e^{i(j-1/2)\varphi } \\
0%
\end{pmatrix}%
,  \label{initstate}
\end{equation}
complete Rabi oscillations appear in the resonant case. That is, the
time dependent quantum mechanical expectation value of the $z$ component of
the spin oscillates between $-1/2$ and $1/2$ (in units of $\hbar )$ at a
given point of the ring:
\begin{equation}
\bar{S}_{z}(\tau ,\varphi )=\bar{S}_{z}(0,\varphi )\cos \Omega _{R}\tau =%
\frac{1}{2}\cos \Omega _{R}\tau.
\end{equation}
(The amplitude of these oscillations are smaller for nonresonant external
fields.) In the time evolution of the $\bar{S}_{x}$ and $\bar{S}_{y}$
expectation values, oscillations with the frequency of the external field
superimpose on the Rabi flopping:
\begin{eqnarray}
\bar{S}_{x}(\tau ,\varphi ) &=&-\frac{1}{2}\sin (\Omega _{R}\tau )\sin (%
\tilde{\nu}\tau -\varphi ) \\
\bar{S}_{y}(\tau ,\varphi ) &=&\frac{1}{2}\sin (\Omega _{R}\tau )\cos (%
\tilde{\nu}\tau -\varphi ).  \notag
\end{eqnarray}%
Fig.~\ref{rabifig} visualizes the spin precession for the resonant case. An
arrow starting from a certain point of the ring points into the direction
of the spin at that spatial position\cite{SC06,KFBP08}. The lengths of the arrows are
proportional to the local electron density regardless of the spin direction:
\begin{equation}
\rho(\varphi ,t)=\langle \Psi (\varphi ,t)|\Psi
(\varphi ,t)\rangle ,
\label{spininner}
\end{equation}%
with $\langle .|.\rangle $ denoting the inner product of spinors (without
integrating over the spatial degrees of freedom.)
Note that in the current case $\rho$ -- and consequently the lengths of the arrows --
do not depend on the position, in this sense the oscillations affect only the spin degrees of freedom.

Let us note that if the initial state is not a pure quantum mechanical
state, e.g., it is an equal weight incoherent sum of the eigenstates (that
is, the spin part of its density operator is proportional to unity), then no
Rabi oscillations will be visible. Therefore in a usual, not specially
prepared sample in equilibrium at low temperatures the question whether these
oscillations can be observed depend on the position of the Fermi level: if
it is between $E_{j}^{+}$ and $E_{j}^{-}$ for some value of $j,$ then Rabi
flopping can be present, while when all occupied states are "paired",
i.e., the eigenstates corresponding to $E_{j}^{+}$ and $E_{j}^{-}$ are both
occupied, their opposite precession cancels out.

\subsection{Time evolution of wave packets}

In contrast to the previous subsection, where we have treated states with
well defined angular momentum, we consider here the time evolution of a
complementary situation, when the initial state is a localized wave
packet with a small uncertainty in the position variable $\varphi .$
Starting with a narrow, spin-polarized initial state, one expects it to
diffuse so that the spin direction changes locally during the process.
However, the ring geometry implies that the "tail" and "head" of the spreading
wave packet will interfere when they start to overlap. This leads to rather
complex dynamics with an additional characteristic feature that can be seen
at this point. Namely, if we assume that there is a finite number of
non-negligible coefficients in the expansion of the initial state in terms
of the eigenstates of the Hamiltonian, the discrete nature of the spectrum
(which is again a consequence of the geometry) may cause the initial phases
to be restored after a certain "revival time". In other words, we expect
periodic "collapse and revival" phenomena: The initially localized wave
packet becomes delocalized along the ring, but later it reassembles again.
(Note that the term "collapse" -- similarly to the case of an atom
interacting with a quantized field\cite{ENS80} -- means here merely the decay of some
expectation values and completely unrelated to the notion of measurement
induced "collapse of the wave function".) In principle, when the
ratio of some important frequencies is irrational (i.e., it is not a
fraction of two integers), the revival time is infinitely long. However,
revivals appear when all frequencies can be written essentially as integer
multiples of a base frequency. This condition can be met by appropriately
choosing the amplitude $\tilde{A}=A/\Omega$ of the SOI oscillations in Eq.~(\ref{omt}) so that $%
1+\tilde{A}^{2}=(m/k)^{2}$ \ with $\mathrm{integer}$ $k$ and $m.$ Then
\begin{equation}
E_{n}^{\pm }\approx (j^{2}+\frac{1}{4})\pm j\frac{m}{k},\ \ k,m\ \mathrm{%
integer},  \label{approxenergies}
\end{equation}%
and the base frequency is $\Omega /k,$ leading to revival times $T_{r}=2k\pi
/\Omega .$ The approximation above is valid if $\tilde{\nu}$ is negligibly
small compared to the relevant values of $j.$ Clearly, this requirement
cannot be met always, but as we will see, rapidly oscillating SOI strengths
modify the above picture only in the sense that revivals become less
pronounced.

\begin{figure}[tbh]
\includegraphics[width=8cm]{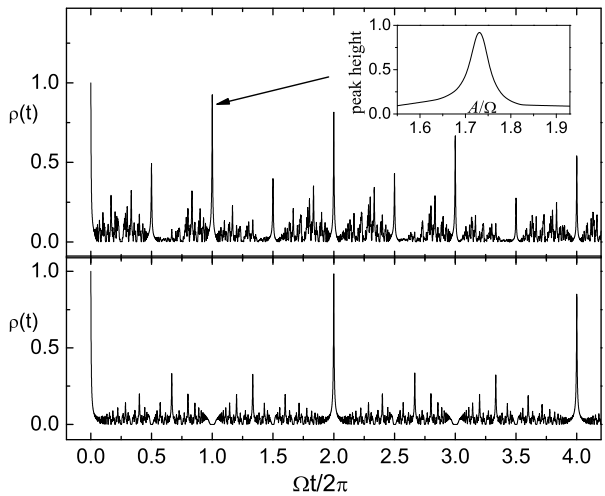}
\caption{The time evolution of the electron density at $\protect\varphi =0$
for the initial state shown in Fig.~\protect\ref{packetfig}. The parameters
are $\tilde{A}=\protect\sqrt{3},\tilde{\protect\nu}=1.0$ (upper panel) and $\tilde{A}=%
\protect\sqrt{21/4},\tilde{\protect\nu}=0.1$ (lower panel); these values
satisfy Eq.~(\protect\ref{approxenergies}) with $m=2,k=1$ and $m=5,k=2,$
respectively. The difference in the periodicity can clearly be seen, and
signatures of partial revivals are also present. The inset of the first panel shows the dependence of the
height of the first revival peak (denoted by the arrow) as a function of $\tilde{A}.$}
\label{densfig}
\end{figure}

\begin{figure}[tbh]
\psfrag{label0}{$\frac{\tau}{2\pi}=0$} \psfrag{label15}{$\frac{\tau}{2\pi}=0.025$} \psfrag{label105}{$\frac{\tau}{2\pi}=0.167$} %
\psfrag{label157}{$\frac{\tau}{2\pi}=0.25$} \psfrag{label314}{$\frac{\tau}{2\pi}=0.5$} \psfrag{label393}{$\frac{\tau}{2\pi}=0.625$} %
\psfrag{label471}{$\frac{\tau}{2\pi}=0.75$} \psfrag{label628}{$\frac{\tau}{2\pi}=1.0$} \includegraphics[width=8cm]{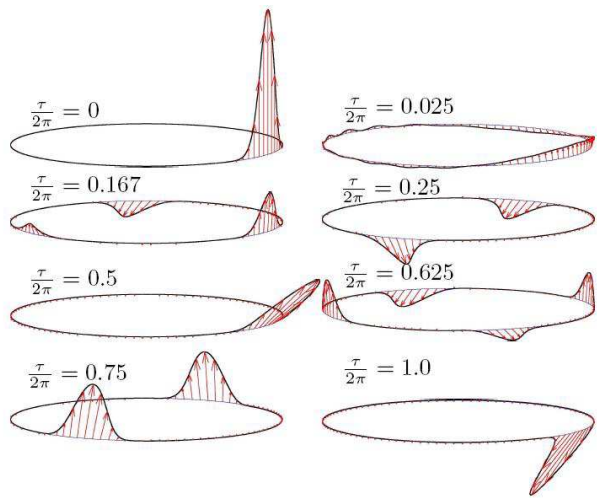}
\caption{Wave packet motion in a quantum ring ($\tilde{A}=\protect\sqrt{3},\tilde{%
\protect\nu}=1.0$.) Note the appearance of "spintronic Schr\"{o}dinger-cat"
states. See EPAPS Document No. [number will be inserted by publisher] for the related movie file.}
\label{packetfig}
\end{figure}

Fig.~\ref{densfig} shows the time evolution of the electron density given
by Eq.~(\ref{spininner}) at $\varphi=0$ for a wave packet which is initially
polarized in the positive $z$ direction and centered at $\varphi=0$ with
Gaussian envelope (see Fig.~\ref{packetfig}.) In the upper (lower) panel $%
\tilde{A}^2=3$ (21/4), Eq.~(\ref{approxenergies}) is satisfied with $m=2,k=1$ $(m=5,
k=2),$ thus the revival time $T_r$ is very close to $2\pi/\Omega$ $%
(4\pi/\Omega).$ The initial phase relations are restored periodically,
although for larger values of $\tilde{\nu}$ the amplitude of the consecutive
revivals decay faster. Clearly, the requirement of commensurable frequencies is related to a rather mathematical point of view, it is hardly possible to exactly satisfy it in an actual experiment. Therefore we investigated less ideal parameters as well, and, according to the inset of Fig.~\ref{densfig}, less pronounced revivals appear also when the parameters are not exactly the ideal ones.

However, an additional, genuinely nonlinear effect can also be seen in Fig.~%
\ref{densfig}: partial revivals at certain time instants $T_{r}/m$, with $m$
being an integer. The significance of these less pronounced peaks can be
seen in Fig.~\ref{packetfig}, where the dynamics of the spinor valued wave
function is visualized in the same way as in Fig.~\ref{rabifig}.
As we can see, the initially localized wave packet becomes first delocalized
(collapse), then, at the partial revival times there is a superposition of
wave packets localized at different positions, and finally, at $T_{r},$ we
can see a single wave packet again (revival). Let us note that the emergence
of the "spintronic Scr\"{o}dinger cat" states is a typical nonlinear
feature, similar phenomena appear e.g., in the case of a wave packet moving
in a Morse potential\cite{FC02}. Finally, let us note that so far we considered only
the effect of the energy levels, but the oscillation of the external field
induces an additional, overall rotation of the spin direction around the $z$
axis at a frequency of $\nu .$ This is why although the spatial form of the
initial wave packets are restored at $T_{r},$ the spin direction is usually
different from the initial one.

\section{Rings with attached leads: conductance properties}

\subsection{Local energy balance}

One of the most direct ways of gaining information of a semiconductor
device is measuring its conductance. Clearly, the model discussed so far
cannot predict properly the result of such experiments, effects related to
the leads that connect the device to the contacts have to be considered as
well. When the Hamiltonian of the ring does not depend on time, this problem
has already been solved \cite{MPV}. In that case basically three
requirements should be met to obtain the solution: i) Energy conservation,
ii) continuity of the wave function, and iii) vanishing net spin current
densities at the junctions. Calculations of the conductance are usually
carried out at a fixed energy (practically at the Fermi energy, as at low
temperatures electrons around this energy level determine conductance
properties), in which case energy conservation is trivially satisfied.
Having solved the eigenvalue problem of the Hamiltonians in the different
domains (ring sections and leads), points ii) and iii) above imply that the
resulting wave functions have to be joined together. The spinor components have to be
continuous, and the spin current \cite{MPV} that enters a junction, should
also leave it:
\begin{equation}
\sum_{{\mathrm{all}}\,{\mathrm{leads}}}\boldsymbol{j}_{s}=0.
\end{equation}%
Let us note that besides the Griffith's boundary conditions\cite{G53} that have been
described above (and will be used throughout this paper) there are other
physically realistic and often used possibilities as well. The choice of the
boundary conditions is in fact shown to be related to the reduction of a two
dimensional problem to one dimension\cite{V09}.

However, in our case, when the Hamiltonian itself contains explicit time
dependence, energy conservation cannot simply be taken into account by solving
the problem within a given eigensubspace of the Hamiltonian. Instead, we use
a continuity equation containing an explicit source term $\mathcal{S}$
\begin{equation}
\frac{\partial }{\partial t}\rho _{E}(\varphi,t)=\nabla \boldsymbol{j}_{E}+%
\mathcal{S}=\frac{\partial }{\partial \varphi }\boldsymbol{j}_{E}(\varphi
,t)+\mathcal{S}(\varphi ,t),  \label{cont}
\end{equation}%
for the local energy density
\begin{equation}
\rho _{E}(\varphi ,t)=\mathrm{Re}\langle \Psi (\varphi ,t)|H(\varphi ,t)\Psi
(\varphi ,t)\rangle ,
\label{edens}
\end{equation}%
with the spinor inner product of Eq.~(\ref{spininner}). Note that, however,
the Hamiltonian (\ref{Ham}) is Hermitian only with respect to the inner product that
involves spatial degrees of freedom as well (see Eq.~(\ref{completeinner})),
that is why we take the real part of the expectation value in Eq.~(\ref{edens}).
Let us note that the time dependent Hamiltonian (\ref{Ham}) can be interpreted to describe the circular motion of a nonrelativistic charged particle in the presence of the scalar potential $\hbar \omega/4e\Omega$ and a spin dependent vector potential. This latter means an effective, spin-dependent electric field\cite{GTF03} being proportional to the time derivative of $\omega(t),$ and can also be related to an effective electromotive force\cite{MTC03} that can induce spin currents.

A calculation similar to that given in the Appendix of Ref.~[\onlinecite{MPV}]
leads to:
\begin{equation}
\mathcal{S}(\varphi ,t)=\mathrm{Re}\langle \Psi (\varphi ,t)|\frac{\partial
}{\partial t}H(\varphi ,t)\Psi (\varphi ,t)\rangle,   \label{energysource}
\end{equation}%
and
\begin{equation}
\begin{aligned}
&\boldsymbol{j}_{E}(t,\varphi ) =  \\
& \ \mathrm{Re}\left( i\Omega (\langle \Psi |\frac{\partial }{\partial \varphi
}H\Psi \rangle -\langle \frac{\partial }{\partial \varphi }\Psi |H\Psi
\rangle )-\omega \langle \Psi |\sigma _{r}H\Psi \rangle \right).
\end{aligned}
\label{energycurrent}
\end{equation}%
The continuity equation (\ref{cont}) is a local relation, its physical
meaning is seen most clearly when it is integrated over a certain domain:
the change of the energy inside the domain is a consequence of the energy
currents that flow in/out through the boundaries, plus the source term
related to the time dependence of the Hamiltonian. In our case this is due
to the oscillating strength of the spin-orbit interaction, so basically the
time dependent electric field between the gate electrodes provides this
extra energy. In the limit when the domain reduces to a single point, e.g.,
to one of the junctions, finite terms (e.g. the source term) that are
integrated over this domain of zero measure disappear, thus Eq.~(\ref{cont})
reduces to
\begin{equation}
\sum_{{\mathrm{all}}\,{\mathrm{leads}}}\boldsymbol{j}_{E}=0,
\label{vanishjE}
\end{equation}%
that is, the net energy current density has to vanish.

\subsection{Solution with time dependent boundary conditions}

To be concrete, let us consider the geometry shown in Fig.~\ref{ringfig} and assume
for the sake of simplicity that the strength of the SOI is zero in the
leads. An incoming spinor valued wave is assumed to reach the device through
lead $I$, and then it is generally partially reflected. Thus using expansion
in terms of plane waves, we may write
\begin{equation}
\left\vert \Psi _{I}\right\rangle =\int_{0}^{\infty }e^{-iE(k)\tau }\left(
e^{ikx}%
\begin{pmatrix}
f_{1}(k) \\
f_{2}(k)%
\end{pmatrix}%
+e^{-ikx}%
\begin{pmatrix}
r_{1}(k) \\
r_{2}(k)%
\end{pmatrix}%
\right) dk,
\end{equation}%
with $E(k)=\frac{\hbar ^{2}k^{2}}{2m^{\ast }\hbar \Omega }=k^{2}a^{2}.$ (Note that e.g. the value of $ka=20.4$ corresponds to a ring with $a=250$ nm at the Fermi energy (11.13 meV) of InGaAs.)

There is no incoming wave in the outgoing arm, that is
\begin{equation}
\left| \Psi_{II} \right\rangle=\int_{0}^{\infty} e^{ikx-iE(k)\tau}
\begin{pmatrix}
t_1(k) \\
t_2(k)%
\end{pmatrix}
d k.
\end{equation}
In the interaction picture introduced by Eq.~(\ref{intpict}), the spinor
components in the incoming and outgoing leads obtain time dependent phases,
e.g, $(f_1(k),f_2(k))$ have to be replaced by $(f_1(k) e^{i\tilde{\nu}%
\tau/2},f_2(k)e^{-i\tilde{\nu}\tau/2}).$

As a consequence of the presence of the leads, the index $j$ of the
eigenstates of the Hamiltonian (\ref{Ham}) does not need to be an integer any more,
and in agreement with the notation used in Ref.~[\onlinecite{FMBP05a}] we replace $j$
by the continuous quantum number $\kappa .$ We expand the wave function in
the two arms using this continuous variable and accordingly shift it from
the index to the argument. In the interaction picture mentioned above we
have
\begin{equation}
\begin{aligned}
\left\vert \Psi _{u}\right\rangle  &=\int_{-\infty }^{\infty }B^{+}(\kappa
)e^{-iE^{+}(\kappa )\tau }%
\begin{pmatrix}
u(\kappa )e^{i(\kappa -1/2)\varphi } \\
v(\kappa )e^{i(\kappa +1/2)\varphi }%
\end{pmatrix}%
d\kappa   \\
&+\int_{-\infty }^{\infty }B^{-}(\kappa )e^{-iE^{-}(\kappa )\tau }%
\begin{pmatrix}
-v(\kappa )e^{i(\kappa -1/2)\varphi } \\
u(\kappa )e^{i(\kappa +1/2)\varphi }%
\end{pmatrix}%
d\kappa.
\end{aligned}
\end{equation}%
A similar expression can be written for the spinor valued wave function in the
lower arm, where the expansion coefficients will be denoted by $C^{+}(\kappa
)$ and $C^{-}(\kappa )$.

In these equations the functions $f_{1}(k)$ and $f_{2}(k)$ are assumed to be
known, and we have to determine $t_{1}(k),t_{2}(k),r_{1}(k),r_{2}(k)$ and $%
B^{\pm }(\kappa ),C^{\pm }(\kappa )$ from the boundary conditions. For the
sake of definiteness let us focus on junction 1. The requirement of
continuity reads:
\begin{equation}
\left\vert \Psi _{I}(x=0,\tau )\right\rangle =\left\vert \Psi _{u}(\varphi
_{1}=0,\tau )\right\rangle =\left\vert \Psi _{l}(\varphi _{2}=0,\tau
)\right\rangle ,  \label{contat1}
\end{equation}%
and it has to be satisfied at all times $\tau $. Provided this holds, it can
be shown that the requirement of vanishing energy current density at this
junction is satisfied, if
\begin{equation}
\begin{aligned} &a\left.\frac{\partial}{\partial x}\left| \Psi_{I} (x,\tau)
\right\rangle\right\vert_{x=0} = \\ &\left.\frac{\partial}{\partial \varphi_1}
\left| \Psi_{u} (\varphi_1,\tau) \right\rangle\right\vert_{\varphi_1=0} +
\left.\frac{\partial}{\partial \varphi_2} \left| \Psi_{l} (\varphi_2,\tau)
\right\rangle\right\vert_{\varphi_2=0}, \label{vanishat1} \end{aligned}
\end{equation}%
similarly to the case of a constant Hamiltonian, but now for arbitrary $\tau.$
In other words, energy conservation follows from the fact that
the requirements ii) and iii) mentioned in the introductory part of this section
are satisfied at any given time.

The set of equations (\ref{contat1}) and (\ref{vanishat1}) together with the
corresponding ones for junction 2 are solved most straightforwardly by
Fourier transformation with respect to $\tau $. E.g, the first equation in (%
\ref{contat1}) is transformed as:
\begin{equation*}
\int \left\vert \Psi _{I}(x=0,\tau )\right\rangle e^{iw\tau }d\tau =\int
\left\vert \Psi _{u}(\varphi _{1}=0,\tau )\right\rangle e^{iw\tau }d\tau
\end{equation*}
which leads to
\begin{eqnarray}
&&%
\begin{pmatrix}
\tilde{f}_{1}(k^{+}(w)) \\
\tilde{f}_{2}(k^{-}(w))%
\end{pmatrix}%
+%
\begin{pmatrix}
\tilde{r}_{1}(k^{+}(w)) \\
\tilde{r}_{2}(k^{-}(w))%
\end{pmatrix}%
= \\
&&\tilde{B}^{+}(\kappa _{1}^{+}(w))|\psi ^{+}(\kappa _{1}^{+}(w))\rangle +%
\tilde{B}^{+}(\kappa _{2}^{+}(w))|\psi ^{+}(\kappa _{2}^{+}(w))\rangle
\notag \\
&&\tilde{B}^{-}(\kappa _{1}^{-}(w))|\psi ^{-}(\kappa _{1}^{-}(w))\rangle +%
\tilde{B}^{-}(\kappa _{2}^{-}(w))|\psi ^{-}(\kappa _{2}^{-}(w))\rangle
\notag
\end{eqnarray}%
where $k^{\pm }(w)=\sqrt{w\mp \tilde{\nu} /2}/a$ \ if the argument of the
square root is not negative, and zero otherwise. Similarly, $\kappa
_{1}^{+}(w)$ and $\kappa _{2}^{+}(w)$ are the two solutions of the equation $%
E^{+}(\kappa )=w$ and similarly $E^{-}(\kappa _{2}^{-}(w))=E^{-}(\kappa _{1}^{-}(w))=w
.$ Additionally,
\begin{equation}
\tilde{B}^{\pm }(\kappa )=\frac{B^{\pm }(\kappa )}{|\partial E^{\pm }(\kappa
)/\partial \kappa |},
\end{equation}%
and $\tilde{f},$ $\tilde{r}$ also denote $f$ and $r$ divided by the
modulus of $\partial E(k)/\partial k.$

For a given value of $w,$ Eqs.~(\ref{contat1}) and (\ref{vanishat1}) and
their counterparts at junction 2 leads to a closed set of 12 linear
equations with 12 unknowns (which are the expansion coefficients $r_{1},r_{2},
$ etc. evaluated at certain values of their respective arguments). This
means that by sweeping $w$ so that $k^{\pm }(w)$ covers the range where $%
f_{1}(k^{+}(w))$ and $f_{2}(k^{-}(w))$ are nonzero, we can obtain all nonzero
values of the unknown functions. Thus inverse Fourier transformation can be
used to calculate the solution of the time dependent transport problem.
(Note that the linearity of the equations implies that when $%
f_{1}(k^{+}(w))=f_{2}(k^{-}(w))=0$, the solution also vanishes everywhere.)
This method provides a general framework to investigate the conductance
properties of the device with oscillating SOI strength.

\subsection{Discussion}

\begin{figure}[tbh]
\includegraphics[width=8cm]{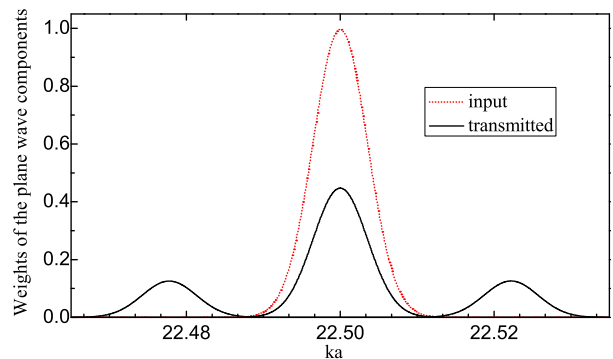}
\caption{The weight $|t_{1}(k)|^{2}+|t_{2}(k)|^{2}$ of the transmitted plane waves as a function of the
wave number $k$ (solid black curve). The incoming distribution $%
|f_{1}(k)|^{2}+|f_{2}(k)|^{2}$ is shown by the dotted line.
The triplet structure is a consequence of scattering events $k\rightarrow
\protect\sqrt{k^{2}\pm \tilde{\protect\nu}/a^{2}}.$ The parameters are $\tilde{A}=0.5, \tilde{\nu}=1.0.$}
\label{tripletfig}
\end{figure}

First we consider the case when the incoming wave has a narrow energy
distribution, as shown by the dotted line in Fig.~\ref{tripletfig}.
For the sake of simplicity we assume a completely unpolarized input
spin state, and calculate the weights of the plane waves $e^{ikx}$ in the
output. If the external electric field were constant, this latter distribution
of the spatial frequencies given by $|t_{1}(k)|^{2}+|t_{2}(k)|^{2}$ would be zero outside the support of $%
|f_{1}(k)|^{2}+|f_{2}(k)|^{2},$ as an incoming plane wave with a given value
of $k$ would lead to transmitted states with the same wave number. (The
linearity of the problem would forbid those values of $k$ in the output, for
which $|f_{1}(k)|^{2}+|f_{2}(k)|^{2}=0.$)  However, as shown in Fig.~%
\ref{tripletfig}, this is not the case, when the SOI strength oscillates.
Besides the central peak corresponding to the wave numbers contained in the
input, there are two additional values of $k$, where $%
|t_{1}(k)|^{2}+|t_{2}(k)|^{2}$ has pronounced maxima. Calculating the
separation of the neighboring peaks in frequency units, we obtain that both
are equal to the frequency of the SOI oscillations, $\nu $.
Mathematically, this result is a consequence of the structure of the fitting
equations, namely that e.g.~$f_{1}(k)$ is connected to both $t_{1}(k)$ \emph{%
and} $t_{2}(\sqrt{k^{2}-\tilde{\nu}/a^{2}}):$
\begin{eqnarray}
f_{2}(k) &\rightarrow &\left\{
\begin{array}{ll}
t_{1}(\sqrt{k^{2}+\tilde{\nu}/a^{2}}) &  \\
t_{2}(k) &
\end{array}%
\right.   \notag \\
f_{1}(k) &\rightarrow &\left\{
\begin{array}{ll}
t_{1}(k) &  \\
t_{2}(\sqrt{k^{2}-\tilde{\nu}/a^{2}}). &
\end{array}%
\right.
\end{eqnarray}%
In a somewhat wider context, we can say that the effect is related to the
time-energy uncertainty relation: when the characteristic time of a process
is not infinitely long, it cannot correspond to a well-defined energy value.
Let us also note that the "sideband currents" shown in Fig.~\ref{tripletfig}
also appear when an oscillating scatterer is being placed in the ring\cite{MB03,CR06}.

Considering the limit of infinitely narrow incoming distribution, i.e, when $%
f_{1}(k)$ and $f_{2}(k)$ are either zero, or proportional to $\delta
(k-k_{0}),$ we may ask what the conductance of the device is. According to
the previous results, in general, the transmitted (and reflected) state
contains wave numbers $k_{0}$ and $\sqrt{k_{0}^{2}\pm \tilde{\nu}/a^{2}}.$ (Note,
however, that at zero temperature, if $k_{0}$ represents the Fermi wave
number, then the output corresponding to $k=\sqrt{k_{0}^{2}-\tilde{\nu}/a^{2}%
}$ is suppressed due to the occupation of levels with $k<k_{0}.$) Then a
direct calculation of the output current shows that generally it will not be
constant, but oscillations with the frequency of the external field appear.
Therefore we calculate average conductance, when time dependent cross terms
disappear due to averaging over a period of $T=2\pi /\nu .$ According to the
Landauer-B\"{u}ttiker formula, this averaged conductance in units of $%
e^{2}/\hbar $ is given by
\begin{equation}
\begin{aligned}
\overline{G}&=\sum \left\vert t_{1}(k_{0})\right\vert ^{2}+\left\vert
t_{2}(k_{0})\right\vert ^{2}\\
&+\frac{\sqrt{k_{0}^{2}+\tilde{\nu}/a^{2}}}{k_{0}}\left\vert
t_{1}(\sqrt{k_{0}^{2}+\tilde{\nu}/a^{2}})\right\vert ^{2}\\
&+\frac{\sqrt{k_{0}^{2}-\tilde{\nu}/a^{2}}}{k_{0}}\left\vert
t_{2}(\sqrt{k_{0}^{2}-\tilde{\nu}/a^{2}})\right\vert ^{2},  \label{LB}
\end{aligned}
\end{equation}
where the sum runs over two orthogonal inputs, e.g.~ $(\exp (ik_{0}x),0)$ and
$(0,\exp (ik_{0}x)).$
\begin{figure}[tbh]
\includegraphics[width=8cm]{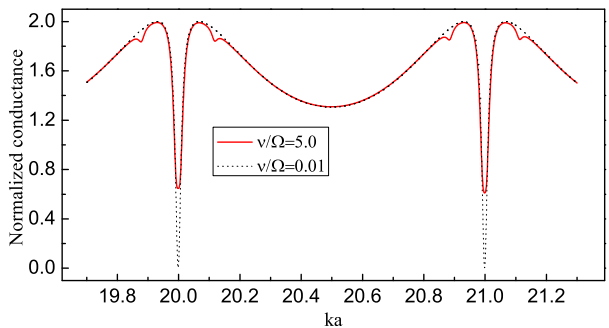}
\caption{Time averaged conductance given by Eq.~(\protect\ref{LB}) for the case of $\tilde{A}=0.5$ as a function of $ka.$}
\label{transfig}
\end{figure}
The result of this calculation is shown in Fig.~\ref{transfig}. As we can
see, the triplet structure shown in Fig.~\ref{tripletfig} appears again,
here in the form of minima in the conductance. Close to the pronounced
minimum of $\overline{G}$ around e.g.~$k_{min}a=20,$ there are two additional
minima corresponding to $k=\sqrt{k_{min}^{2}\pm\tilde{\nu} a^{2}}.$ This `hole
burning' in the transmission is the consequence of the fact that the minima of the
third and fourth terms in Eq.~(\ref{LB}) are situated at different positions from
those of the first two terms. In other words, scattering events that
change the momentum, also modify the conductance of the device. \bigskip

\begin{figure}[tbh]
\psfrag{t0}{$\tau=0$} \psfrag{t4}{$\tau=0.1$} %
\psfrag{t6}{$\tau=0.15$} \psfrag{t8}{$\tau=0.2$}
\includegraphics[width=8cm]{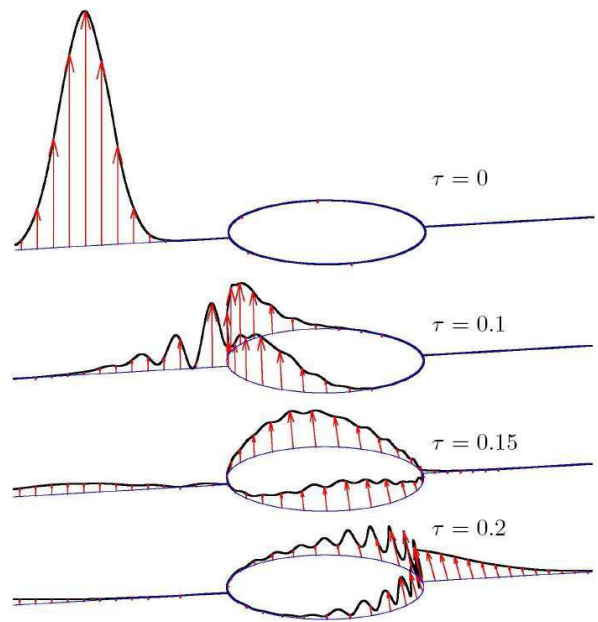}
\caption{Transmission of a packet through the quantum ring ($\tilde{A}=0.5, \tilde{\nu}=0.5$). The initial state is given by Eq.~(\protect\ref{packetinput}), with $\sigma=0.5a, x_0=1.5 a.$ The round trip time is given by $\tau_{\text{round}}=\pi/(k_0a)=0.2,$ where $k_0a=10.$ See EPAPS Document No. [number will be inserted by publisher] for the related movie file.}
\label{ptransfig}
\end{figure}

Finally let us investigate the transmission of a wave packet through the
ring. We choose here the case which in some sense is the opposite to the one
discussed above, as now the distribution $|f_{1}(k)|^{2}+|f_{2}(k)|^{2}$ is
wide in $k$, resulting in a narrow wave packet in space. For the sake of
definiteness, the initial state is assumed to be spin-polarized in the
positive $z$ direction, and localized in the input arm with Gaussian
envelope
\begin{equation}
\left\vert \Psi (x,t=0)\right\rangle =%
\begin{pmatrix}
e^{-\frac{(x-x_{0})^{2}}{\sigma^2}+ik_{0}x} \\
0%
\end{pmatrix}%
\label{packetinput}
\end{equation}%
as shown in Fig.~\ref{ptransfig} for the case of $\sigma=0.5a,$ $x_{0}=1.5a$ and $k_{0}a=10.$ As we can see,
when the wave packet reaches the ring, it is partially reflected, and the
reflected waves interfere with the incoming packet, that leads to
oscillating electron density in the input arm. The fraction of the wave
packet that enters the ring travels along the two arms, interferes around
the output junction and produces an output wave packet. However, there is a
certain probability for the electron not to leave the ring, there is a
fraction of the wave packet that moves backwards toward the input junction,
where interference phenomena can be observed again, and a weak reflected wave
packet is formed that leaves the ring through the input junction. These
processes are repeated periodically until the modulus of the wave function
becomes negligible. As a consequence, both in transmission and reflection we
can observe a series of wave packets with decreasing amplitudes and
increasing widths. The separation of these wave packets in
time is essentially the round trip time of the wave packet in the ring,
which is around $\pi/(ak_0\Omega)$. Effects related to the SOI are twofold: first the
transmission probability (that determines the heights of the consecutive
transmitted wave packets) depends on the amplitude and frequency of the SOI
oscillations, and the spin directions in the ring are also determined by
these parameters.

\section{Random scatterers, thermal fluctuations and an application}
In this section we investigate to what extent the effects discussed so far are still present in a more realistic context, when random scattering events and thermal fluctuations are also taken into account. To this end, first we introduce point-like random scatterers represented by Dirac-delta potentials. That is, we add a term $U(D)=u(D)\delta(\phi)$ to the Hamiltonian (\ref{Ham}), and solve the time dependent problem determined by this perturbed Hamiltonian. The strength of the potential, $U(D),$ is random, it is drawn from a normal distributions, with zero mean and root-mean-square deviation $D.$ The random fluctuation of the potential causes that the "average" state of the system cannot be represented by a pure quantum mechanical state, it becomes a mixture that can be described by a density operator $\rho(D)$. In practice, $\rho(D)$ is calculated from several computational runs, with appropriate averaging. That is, for a particular value of the potential $U(D),$ we obtain a solution spinor, that we write symbolically as $\left|\Psi[U(D)]\right\rangle.$ When after $M_c$ computational runs, the estimated density operator
\begin{equation}
\rho_{out}(D)=
\frac{1}{M_c} \sum_n  \left|\Psi[U(D)]\right\rangle
\left\langle\Psi[U(D)]\right| \label{rhoscatt},
\end{equation}
converges, we have all the possible information needed to describe what effects result from the disturbances characterized by the variable $D.$  In this way, by tuning $D$ we can model weak disturbances (small $D$) as well as the case when frequent scattering events completely change the character of the transport process (corresponding to large values of $D$).

The energy distribution of the input electrons can be taken into account by appropriate averaging over the possible input energies. In thermal equilibrium at temperature $T$, the conductance of the device can be written as
\begin{equation}
G(T)=\int p(E,T) \overline{G}(E) dE.
\label{rhoT}
\end{equation}
where $p(E,T)=-\frac{\partial}{\partial E}[1+\exp{(E-E_F)/k_B T}]^{-1}$ and $\overline{G}(E)$ is given by Eq.~(\ref{LB}). (Note that this is essentially the Landauer-B\"{u}ttiker formula  at finite temperature and low bias\cite{D95}.) For the sake of numerical convenience, we can convert the integral (\ref{rhoT}) to a sum over the possible energies, meaning that the expression for $\rho_{out}(T)$ is similar to Eq.~(\ref{rhoscatt}), but the weights of the projectors are not uniform, they are determined by the Fermi distribution.

Considering first a closed ring, we found that an individual scatterer shifts the energy levels according to the strength of the Dirac-delta potential. Consequently, when averaging over numerous random scattering events, we obtain a certain broadening of the possible energies as well as the Rabi frequencies. Investigating the time evolution of the expectation value $\bar{S}_{z},$ we found that for the parameters and initial state shown in Fig.~\ref{rabifig}, the height of the first maximum at $t=2\pi\Omega_R$ decreases when we increase $D,$ the width of the random distribution. As a reference, we considered the case of time independent Hamiltonian (when $\omega(t)$ is constant in Eq.~(\ref{Ham})) and calculated the conductance Aharonov-Casher (AC) oscillations for various values of $D.$ We found that the visibility
\begin{equation}
I(D)=\frac{G_{max}(D)-G_{min}(D)}{G_{max}(D)+G_{min}(D)}
\end{equation}
of these experimentally detectable oscillations decreases from unity to $1/2$ while $D$ increases from zero to $D_{1/2}=0.03\times E_F$. ($E_F$ denotes the Fermi energy.) At $D_{1/2},$ the first maximum of $\bar{S}_{z}$ is approximately 50\% of the ideal value, thus a damped oscillation can be detected. On the other hand, the revival phenomena shown in  Figs.~\ref{densfig} and \ref{packetfig} are more sensitive to the broadening of the frequencies and consequently they are not expected to be detectable in usual samples. Note that this is in accordance with the expectations, as the
"spintronic Schr\"odinger-cat sates" shown in Fig.~\ref{packetfig} are highly nonclassical, and consequently they are exceptionally sensitive to any kind of environmental noise\cite{GJK96}.
\bigskip

\begin{figure}[tbh]
\includegraphics[width=8cm]{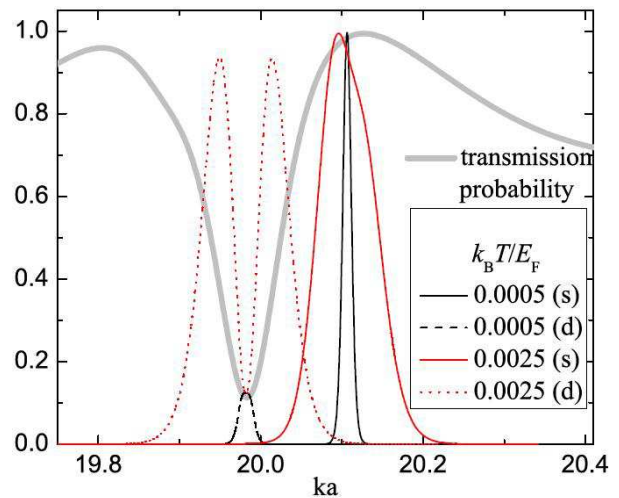}
\caption{Direct (d) and sideband (s) currents when the input energy distribution (centered at Fermi energy corresponding to $ka=19.98$) is broadened due to thermal fluctuations. The relevant part of the transmission profile seen for a wider interval in Fig.~\ref{transfig} is also shown. In this case (parameters are: $\tilde{A}=0.75, \tilde{\nu}=5.0$) the sideband current can be considerably stronger than the direct one. The input spins are chosen to be polarized in the positive $z$ direction and the transmission peaks has been normalized so that the maximum of the higher one is 1, for both temperatures, separately.}
\label{strongsidefig}
\end{figure}

The question to what extent the transport properties of the time dependent problem are modified by random scatterers and thermal fluctuations will be analyzed using a possible application of our model. First we recall Figs.~\ref{tripletfig} and \ref{transfig}, showing a triplet structure in the transmission and the transmission probabilities for infinitely narrow incoming energy distribution. Clearly, the first effect is rather common when oscillating potentials are considered, but the second one is based on the specific spin dependent interference phenomena that appear in a ring structure. Combining these two effects, one can see that it is possible to find parameter values, where the ratio of the transmitted "direct" and "sideband" currents is strongly modified by the transmission profile. Particularly, if the transmission probability for the central peak is suppressed by interferences (like around integer values of $ka$ in Fig.~\ref{transfig}), sideband currents can provide stronger output signals than the direct one. (Usually, in other physical systems, it is mainly the strength of the oscillating potential that determines the relative intensity of the relevant frequencies in the output.) This situation is shown in Fig.~\ref{strongsidefig} for the case of spin-up input in the $z$ direction. Let us recall that while in the direct current this spin direction remains unchanged, the sideband current corresponds to oppositely polarized output spins.  (Note additionally, that the positions of the minima of the transmission probabilities as a function of $ka$ can be tuned by adding a constant SOI term to the Hamiltonian (\ref{Ham}), a minimum can even be transformed into a maximum by using experimentally achievable SOI strengths\cite{MPV}.) The question whether the two peaks are distinguishable, is clearly related to ratio of $\hbar \nu$ and the width of the thermally broadened input energy distribution: the shift induced by the oscillating SOI strength has to be larger than the width of the distribution.

\begin{figure}[tbh]
\includegraphics[width=8cm]{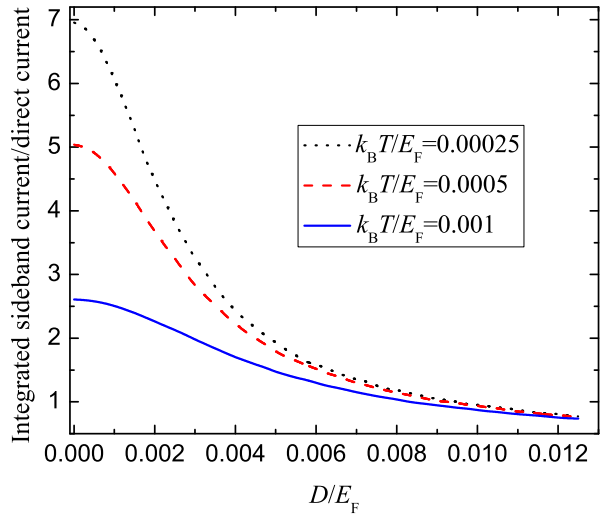}
\caption{The ratio of the integrated  sideband and direct peaks shown in Fig.~\ref{strongsidefig} as a function of the strength of the random scattering events. Integration was carried out with respect to the energy.}
\label{sideintfig}
\end{figure}

Fig.~\ref{strongsidefig} visualizes the case of low temperatures, and does not take random scattering events into account. Now we investigate whether the effects is still visible when $D\neq0$ and the temperature is increased. Choosing experimentally achievable temperatures\cite{KTHS06}, Fig.~\ref{sideintfig} shows the ratio $\int I_{s}(E) dE / \int I_{d}(E) dE$ of the sideband and direct peaks (seen in Fig.~\ref{strongsidefig}) as a function of the strength $D$ of the random scattering events. (Numerically, if $E_F$ is assumed to be in the range of 10 meV, the temperatures corresponding to the curves shown in Fig.~\ref{sideintfig} have the order of magnitude of 100 mK.) As we can see, the effect of pronounced sideband peaks is still present for moderate values of $D.$

Finally let us note that there are samples where the transport is due to many channels in the ring, thus our results are not directly applicable. For narrow rings, however, the analysis of this section shows that some of the previously predicted effects -- that are based on an idealized description -- are stable enough against random scatterers and thermal fluctuations to be visible.

\section{Summary}

In this paper we investigated quantum rings with time dependent spin-orbit
interaction (SOI). The effect of the sinusoidal SOI strength was first investigated
in an isolated ring, i.e., a ring without any attached leads.
In this case we have shown that for initial states with
well-defined $z$ component of the total angular momentum Rabi oscillations
appear: The spin direction along the ring changes periodically in time.
Considering the dynamics of an initially localized wave packet, we
demonstrated that appropriately chosen amplitudes of the SOI oscillations
can lead to quasi-periodic time evolution: First the wave packet becomes
delocalized (this process can be termed as collapse), then at the revival
time it reassembles again. During this process (without thermal fluctuation or scattering events), there are time instants (partial revival times) when "spintronic Schr\"{o}dinger-cat states" appear:
superpositions of wave packets localized at different positions along the
ring.

In the second part of the paper we focused on the transmission properties of
the device. We introduced a general method to treat the quantum
mechanical scattering problem for a time dependent Hamiltonian. Based on
this result, we have shown that in general the scattered outgoing plane wave
states will have an energy distribution which is different from that of the
incoming states. For a single incoming plane wave with energy $E,$ the
output shall contain energy values $E$ and $E\pm \hbar \nu $, that is,
shifts corresponding to the frequency of the SOI oscillation appear.
If the input is spinpolarized e.g.~in the positive $z$ direction, the sideband current corresponding to $E+\hbar\nu$ is related to oppositely polarized output. Additionally, it was shown that one can find parameter values, where sideband currents can provide stronger output signal than the direct one. This transport property is specific to quantum rings: sideband currents appear in a wide class of physical systems, but the most important role in the  suppression of the direct current is played by spin dependent quantum interference, which is closely related to the geometry of the device. Additionally, this effect is still visible at finite temperatures even if random scattering events modify the the dynamics. According to our calculations, Rabi oscillations can also survive a moderate level of temperature induced fluctuations.

\section*{Acknowledgments}

This work was supported by the
Flemish Science Foundation (FWO-Vl), the Belgian Science Policy (IAP) and the
Hungarian Scientific Research Fund (OTKA) under Contracts Nos.~T48888,
M36803, M045596. P.F.~was supported by a J.~Bolyai grant of the Hungarian
Academy of Sciences.

\end{document}